\documentclass[conference]{IEEEtran}
\usepackage[T1]{fontenc}
\usepackage[utf8]{inputenc}
\usepackage{amssymb}
\usepackage[pdftex]{graphicx}
\usepackage{dsfont}
\usepackage{siunitx}
\usepackage{mathtools}
\usepackage{mathabx}
\usepackage{array}
\usepackage[noadjust]{cite}
\usepackage{glossaries}
\usepackage{algorithm}
\usepackage{algpseudocode}
\usepackage{xcolor}
\usepackage{tikz}
\usepackage{pgfplots}
\usepgfplotslibrary{fillbetween}

\usetikzlibrary{decorations.pathreplacing,calc}
\newcommand{\tikzmark}[1]{\tikz[overlay,remember picture] \node (#1) {};}

\newcommand*{\AddNote}[4]{%
    \begin{tikzpicture}[overlay, remember picture]
        \draw [decoration={brace,amplitude=0.5em},decorate]
            ($(#3)!(#1.north)!($(#3)-(0,1)$)$) --  
            ($(#3)!(#2.south)!($(#3)-(0,1)$)$)
                node [align=center, text width=2.5cm, pos=0.5, anchor=west] {#4};
    \end{tikzpicture}
}%

\makeatletter
\let\MYcaption\@makecaption
\makeatother

\usepackage[font=footnotesize]{subcaption}

\makeatletter
\let\@makecaption\MYcaption
\makeatother
\algnewcommand{\LineComment}[1]{\(\triangleright\) #1}

\newacronym{ACM}{ACM}{adaptive coding and modulation}
\newacronym{ADC}{ADC}{analog-to-digital conversion}
\newacronym{AGC}{AGC}{automatic gain control}
\newacronym{AWGN}{AWGN}{additive white Gaussian noise}
\newacronym{BER}{BER}{bit error rate}
\newacronym{BLER}{BLER}{block error rate}
\newacronym{BP}{BP}{backpropagation}
\newacronym{BPTT}{BPTT}{backpropagation through time}
\newacronym{CE}{CE}{cross-entropy}
\newacronym{CFO}{CFO}{carrier frequency offset}
\newacronym{CSI}{CSI}{channel state information}
\newacronym{DAC}{DAC}{digital-to-analog conversion}
\newacronym{DL}{DL}{deep learning}
\newacronym{DFT}{DFT}{discrete Fourier transform}
\newacronym{FFT}{FFT}{fast Fourier transform}
\newacronym{GAN}{GAN}{generative adversarial network}
\newacronym{GRU}{GRU}{gated recurrent unit}
\newacronym{iid}{i.i.d.\@}{independent and identically distributed}
\newacronym{IFFT}{IFFT}{inverse fast Fourier transform}
\newacronym{KL}{KL}{Kullback-Leibler}
\newacronym{LSTM}{LSTM}{long short-term memory}
\newacronym{MDP}{MDP}{Markov decision process}
\newacronym{ML}{ML}{machine learning}
\newacronym{MLP}{MLP}{multilayer perceptron}
\newacronym{MIMO}{MIMO}{multiple-input multiple-output}
\newacronym{MSE}{MSE}{mean squared error}
\newacronym{NN}{NN}{neural network}
\newacronym{OFDM}{OFDM}{orthogonal frequency-division multiplexing}
\newacronym{pdf}{pdf}{probability density function}
\newacronym{pmf}{pmf}{probability mass function}
\newacronym{QPSK}{QPSK}{quadrature phase-shift keying}
\newacronym{PSNR}{PSNR}{Peak Signal to Noise Ratio}
\newacronym{RBF}{RBF}{Rayleigh block-fading}
\newacronym{ReLU}{ReLU}{rectified linear unit}
\newacronym{RTN}{RTN}{radio transformer network}
\newacronym{RL}{RL}{reinforcement learning}
\newacronym{RNN}{RNN}{recurrent neural network}
\newacronym{SFO}{SFO}{sampling frequency offset}
\newacronym{SNR}{SNR}{signal-to-noise ratio}
\newacronym{SINR}{SINR}{signal-to-interference-plus-noise ratio}
\newacronym{SGD}{SGD}{stochastic gradient descent}
\newacronym{wrt}{w.r.t.\@}{with respect to}

\renewcommand{\vec}[1]{\mathbf{#1}}
\newcommand{\vecs}[1]{\boldsymbol{#1}}

\newcommand{\lv}{\vec{l}}
\newcommand{\mv}{\vec{m}}

\newcommand{\pv}{\vec{p}}

\newcommand{\rv}{\vec{r}}

\newcommand{\wv}{\vec{w}}
\newcommand{\xv}{\vec{x}}
\newcommand{\yv}{\vec{y}}

\newcommand{\varepsilonv}{\vecs{\varepsilon}}

\newcommand{\thetav}{\vecs{\theta}}

\newcommand{\Id}{\vec{I}}

\newcommand{\Wm}{\vec{W}}
\newcommand{\Xm}{\vec{X}}
\newcommand{\Ym}{\vec{Y}}

\newcommand{\Cc}{{\cal C}}

\newcommand{\Nc}{{\cal N}}

\newcommand{\Uc}{{\cal U}}

\newcommand{\CC}{\mathbb{C}}
\newcommand{\MM}{\mathbb{M}}

\newcommand{\RR}{\mathbb{R}}

\newcommand{\tp}{^{\mathsf{T}}}

\newcommand{\LB}{\left(}
\newcommand{\RB}{\right)}
\newcommand{\LP}{\left\{}
\newcommand{\RP}{\right\}}
\newcommand{\LSB}{\left[}
\newcommand{\RSB}{\right]}

\renewcommand{\log}[1]{\mathop{\mathrm{log}}\LB #1\RB}

\newcommand{\EE}{{\mathbb{E}}}

\newcommand\norm[1]{\left\lVert#1\right\rVert}

\begin{document}
\title{Deep Reinforcement Learning Autoencoder with Noisy Feedback}
\author{
\IEEEauthorblockN{Mathieu Goutay, Fayçal Ait Aoudia, and Jakob Hoydis}
\IEEEauthorblockA{Nokia Bell Labs, Paris-Saclay, 91620 Nozay, France\\ mathieu.goutay@insa-lyon.fr, \{faycal.ait\_aoudia, jakob.hoydis\}@nokia-bell-labs.com
}}
\maketitle

\begin{abstract}
End-to-end learning of communication systems enables joint optimization of transmitter and receiver, implemented as deep \gls{NN}-based autoencoders, over any type of channel and for an arbitrary performance metric.
Recently, an \emph{alternating training} procedure was proposed which eliminates the need for an explicit channel model. However, this approach requires feedback of real-valued losses from the receiver to the transmitter during training.
In this paper, we first show that alternating training works even with a noisy feedback channel. Then, we design a system that learns to transmit real numbers over an unknown channel without a preexisting feedback link.
Once trained, this \emph{feedback system} can be used to communicate losses during alternating training of autoencoders.
Evaluations over \gls{AWGN} and \gls{RBF} channels show that end-to-end communication systems trained using the proposed feedback system achieve the same performance as when trained with a perfect feedback link.
\end{abstract}

\glsresetall

\section{Introduction} \label{sec:intro}

For the last decades, engineers have built communication systems by splitting the transmitter and receiver into multiple components, each optimized for a specific task. This modular block-structure is the basis for the robust and versatile communication systems we have today. Since the early days, \gls{ML} techniques were considered for specific problems in communication \cite{widrow1960adaptive}. For various reasons, however, they have never become the de facto solution and were rarely implemented in commercial products.

\begin{figure}
\centering
	\includegraphics[width=\linewidth]{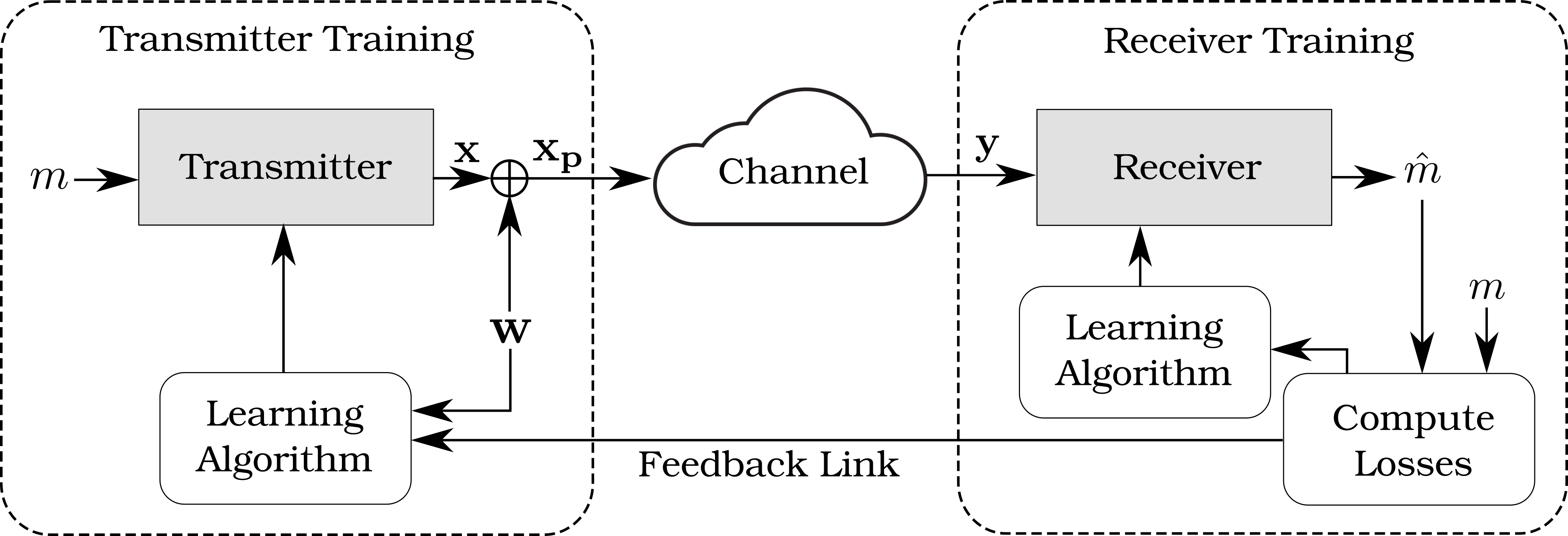}
	\caption{Training of a communication system using a feedback link}
\label{fig:main_plot}
\vspace{-15pt}
\end{figure}

Recently, it was proposed~\cite{8054694} to interpret end-to-end communication systems as \textit{autoencoders}~\cite{Goodfellow-et-al-2016-Book}, where the transmitter and receiver are implemented as deep \glspl{NN}.
The main practical drawback of this theoretically very appealing idea is the need for a mathematical channel model to perform the training. This makes its application to real channels of practical interest challenging.
A first approach to circumvent this problem \cite{dorner2017deep} consists in training the system using a channel model, and then fine-tuning the receiver with measured data. A shortcoming of this approach is  sub-optimal training of the transmitter which makes it not fully satisfactory.
Another solution studied by multiple authors \cite{o2018approximating, ye2018channel} consist in  learning a differentiable channel model in the form of a \gls{GAN}, which can then be used for supervised autoencoder training. However, it still needs to be shown that this approach works for practical channels.
Lastly, the authors of \cite{Aoudia2018EndtoEndLO} proposed a third approach called \emph{alternating training}, where the autoencoder is trained by alternating supervised learning of the receiver, and \gls{RL} of the transmitter. 
A similar method is used in \cite{devrieze2018multiagent}, but the receiver is not actually trained and simply detects symbols through clustering. Another model-free approach is developed in \cite{raj2018backpropagating} based on simultaneous perturbation methods. 
Although all of these methods do not require any knowledge of the channel and can be performed directly with real hardware, a reliable feedback link is needed during training, from the receiver to the transmitter, as illustrated in Fig.~\ref{fig:main_plot}. 

In this work, we show that alternating training of end-to-end communication systems can be performed with noisy feedback, up to a certain point, without performance loss. Based on this observation, we design a system which learns to communicate real values over an unknown channel and without the need for any reliable link.
Once trained, this \emph{feedback system} can be used in lieu of the perfect feedback link assumed in \cite{Aoudia2018EndtoEndLO}.
Therefore, the contribution of this paper, combined with the alternating training algorithm of~\cite{Aoudia2018EndtoEndLO}, enables training of end-to-end communication systems without channel model and without an extra feedback link.
Evaluation of the proposed scheme on \gls{AWGN} and \gls{RBF} channels shows that the proposed approach achieves the same performance as training with perfect feedback. Moreover, the learned system outperforms a coherent \gls{QPSK} baseline as well as a highly optimized higher-order sphere packing scheme \cite{Agrell16} on the \gls{RBF} channel.

\textbf{Notations :} Matrices and column vectors are denoted by bold upper- and lower-case letters, respectively. $x^{(i)}$ is the $i^{\text{th}}$ element of $\xv$.
$||\Xm||_F$  denotes the Frobenius norm of $\Xm$.
$\LSB\xv\RSB^{\bot b}_{\top a}$ is the restriction of the elements of $\xv$ in the interval $[a,b]$ by clipping.
$\Re(\xv)$ and $\Im(\xv)$ denote the real and the imaginary part of $\xv$, respectively. $\mathds{1}_{N}$ is the $N$-dimensional all-one vector and $\Id_N$ the $N$--by--$N$ identity matrix.
The gradient and Jacobian operators with respect to a vector $\thetav$ are both denoted $\nabla_{\thetav}$.

\section{Learning of a Communication System} \label{sec:comm_system}

In an \gls{ML}-based end-to-end communication system, the transmitter and the receiver are both implemented as deep \glspl{NN}, with parameters (weights, bias) $\thetav_T$ and $\thetav_R$, respectively.
The transmitter $f^{T}_{\thetav_T}: \MM \to \CC^{N_c}$  maps a message $m$ uniformly drawn from a finite discrete set $\MM = \LP 1, \dots, M \RP$ to a vector of complex symbols $\xv \in \CC^{N_c}$, which are normalized and sent over the channel. $N_c$ is the number of (complex) channel uses.
The receiver $f^{R}_{\thetav_R} : \CC^{N_c} \to \LP \pv \in \RR_+^M~|~\sum_{i=1}^M p^{(i)} = 1 \RP$ generates a probability distribution over $\MM$ from the received signal $\yv$, and hard decoding is achieved by taking the message $\hat{m}$ with highest probability.
The channel acts as a stochastic black box with unknown transfer function which alters the sent signal $\xv$ according to the conditional distribution $P(\yv|\xv)$.

The key idea of alternating training \cite{Aoudia2018EndtoEndLO} is to find the best sets of parameters by alternating between supervised learning of the receiver and \gls{RL}-based learning of the transmitter (Fig.~\ref{fig:main_plot}).
The aim of receiver training is to learn an estimate of the conditional probability $P(m|\yv)$ and is performed by \gls{SGD}~\cite{Goodfellow-et-al-2016-Book} on the \gls{CE}
\begin{equation}\label{eq:ce-loss}
  L(\thetav_R) = \frac{1}{S_c} \sum_{i=1}^{S_c} \underbrace{-\log{\LSB f_{\thetav_R}^{R}(\yv^{(i)})\RSB^{\LB m^{(i)} \RB}}}_{l^{(i)}} 
\end{equation}
where $S_c$ is the \emph{minibatch} size, i.e., the number of examples $m^{(i)}$ used to estimate the loss $L$, $l^{(i)}$ is the \emph{per example loss}, $\yv^{(i)}$ the signal received when $m^{(i)}$ was sent, and $\LSB f_{\thetav_R}^{R}(\yv^{(i)})\RSB$  the probability distribution vector over all possible messages outputted by the receiver. It is assumed that the examples are \gls{iid}.

Transmitter training is achieved by adding a zero-mean stochastic perturbation $\wv$ to the encoded message, i.e., $\xv_p = \xv + \wv$ denotes the resulting signal transmitted over the channel.
The perturbation is chosen to be Gaussian \gls{iid} with a variance $\sigma_c$, $\wv \thicksim \Cc\Nc(\vec{0}, \sigma_{c}^2\Id_{N_c})$, so the conditional distribution of $\xv_p$ is denoted by $\pi_{\thetav_T}(\xv_p|m)=\Cc\Nc(f^{T}_{\thetav_T} (m), \sigma_{c}^2\Id_{N_c})$.
 The receiver computes the per-example losses $l^{(i)}$ for all  transmitted messages $m^{(i)}$, and sends them back to the transmitter, which optimizes its parameters $\thetav_T$ by \gls{SGD}. The gradient of the aggregate loss $J(\mv, \lv, \Xm_p)$ under the \emph{policy} $\pi_{\thetav_T}$ is estimated using the policy gradient theorem~\cite{NIPS1999_1713} from the field of \gls{RL} :
 
\vspace{-10pt}
\small
\begin{align}
\label{eq:grad_est}
&\nabla_{\thetav_{T}} J(\mv, \lv, \Xm_p) = \frac{1}{S_c} \sum_{i=1}^{S_c} l^{(i)} \underbrace{\nabla_{\thetav_{T}} \log{\pi_{\thetav_T}(\xv_p^{(i)}|m^{(i)}) }}_{\triangleq\  D^{(i)}} \\ 
& = \frac{1}{S_c} \sum_{i=1}^{S_c} \frac{2 l^{(i)}}{\sigma_{c}^2} \LB\nabla_{\thetav_T} f^{T}_{\thetav_T}\LB m^{(i)} \RB \RB\tp\LB \xv_p^{(i)} - f^{T}_{\thetav_T}\LB m^{(i)}\RB \RB \nonumber
\end{align}
\normalsize
where $\lv$ is the vector of per-example losses $l^{(i)}$ and $\Xm_p$ the matrix of perturbed vectors $\xv_p^{(i)}$.

In~\cite{Aoudia2018EndtoEndLO}, the authors assumed that a perfectly reliable link is available to feed back the losses computed by the receiver to the transmitter, which is hardly the case in practice.
In the next section, we study the effect of \emph{noisy} feedback on the end-to-end system training process.

\section{Effect of Noisy Losses on Training} 
\label{sec:effect_noisy_losses}


Assuming a noisy scalar feedback channel, we denote by $\varepsilonv$ the error introduced by the channel, such that $\widetilde{\lv}  = \lv + \varepsilonv$, where $\lv$ denotes the true loss vector, and $\widetilde{\lv}$ the noisy loss.
We model the feedback channel as being \gls{AWGN}, i.e., $\varepsilonv \thicksim \Nc(\vec{0}, \sigma_l^2\Id_{S_c})$.
Note that $\sigma_l^2$ is the \gls{MSE} between $\lv$ and $\widetilde{\lv}$.
We denote by $\nabla_{\thetav_T} \widetilde{J}$ the loss function gradient estimator obtained by replacing the true losses by noisy ones in (\ref{eq:grad_est}).
All the numerical evaluations of this section were performed with $S_c = 10^5$, $M=256$, $N_c = 4$, and $\sigma_{c}^2 = 0.02$.

\subsection{Effect of the Noise Variance on the Loss Function Gradient}

As the error $\varepsilonv$ is assumed to have zero-mean, one can easily see that the loss function gradient estimator is not biased, i.e., $\EE \LSB \nabla_{\thetav_T} \widetilde{J} \RSB = \EE \LSB \nabla_{\thetav_T} J \RSB$.
However, noisy feedback increases the estimator variance, interfering with the communication system training. To get insight into this issue, one can compute the cumulative element-wise variance of $\nabla_{\thetav_T} \widetilde{J}$, denoted by $V$:

\begin{align}\nonumber
 V & \triangleq \EE \LSB \norm{ \nabla_{\thetav_T} \widetilde{J} - \EE \LSB \nabla_{\thetav_T} J \RSB }_2^2 \RSB \\ 
 & = \EE \LSB \norm{ \frac{1}{S_c} \sum_{i=0}^{S_c} \widetilde{l}^{(i)} D^{(i)} - \EE \LSB  \frac{1}{S_c} \sum_{i=0}^{S_c} l^{(i)} D^{(i)} \RSB }_2^2 \RSB \\
& = \frac{1}{S_c} \LB \underbrace{\EE \LSB || l D - \EE \LSB l D \RSB ||_2^2 \RSB}_{\triangleq A} + \underbrace{\sigma_l^2 \EE \LSB ||D||_2^2 \RSB }_{\triangleq B}\RB \nonumber
\end{align}
where $l$ and $D$ refer to the random variables generating the samples $l^{(i)}$ and $D^{(i)}$ (as defined in \eqref{eq:grad_est}), respectively.
It can be noticed that $A$ is the cumulative element-wise variance of the loss function gradient estimator without noise $\nabla_{\thetav_{T}} J$.
Having noisy feedback therefore increases $V$ by the term $B$ which can be written as:
\begin{equation}
B = \frac{4\sigma_l^2}{\sigma_{c}^4} \EE \LSB  || \nabla_{\theta_{T}} f_{\theta_T}^{T}(m) ||_F^2 \RSB.
\end{equation}

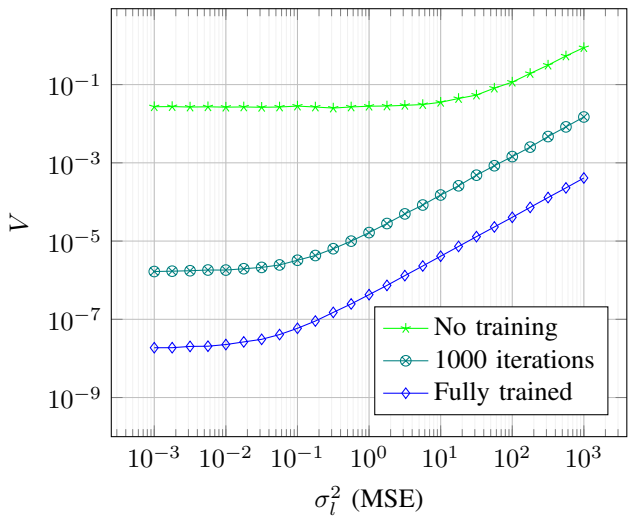
\begin{figure}[!t]
\begin{tikzpicture}
	  \begin{axis}[
	    xmode=log,
	    ymode=log,
	    grid=both,
	    grid style={line width=.01pt, draw=gray!10},
	    major grid style={line width=.2pt,draw=gray!50},
	    minor tick num=1,
	    xlabel={$\sigma_l^2$ (MSE)},
	    ylabel={$V$},
	    	legend style={at={(0.51, 0.05)},anchor=south west},
	    ymin = 1e-10,
	    legend cell align={left},
	  ]
	    \addplot[green, mark=star] table [x=MSE, y=0, col sep=comma] {figs/Var_MSE_final.csv};
	    \addplot[teal, mark=otimes] table [x=MSE, y=1000, col sep=comma] {figs/Var_MSE_final.csv};
		\addplot[blue, mark=diamond] table [x=MSE, y=final, col sep=comma] {figs/Var_MSE_final.csv};
	  \addlegendentry{No training}
	  \addlegendentry{1000 iterations}
	  \addlegendentry{Fully trained}
		\end{axis}
		\end{tikzpicture}
		\vspace{-10pt}
	    \caption{Numerical evaluation of the cumulative element-wise variance of the loss function gradient estimator $V$ for different values of $\sigma_l^2$ (MSE)}
	    \label{fig:Var_MSE}
	    \vspace{-10pt}
\end{figure}

Fig.~\ref{fig:Var_MSE} shows a numerical evaluation of $V$ as a function of $\sigma_l^2$ for an untrained system, after 1000 training iterations, and for a system trained until no significant progress could be observed.
In the three considered cases, $V$ is not significantly impacted by increasing values of $\sigma_l^2$ up to a certain point, from which $V$ increases linearly with $\sigma_l^2$.
From the above analysis, this phenomenon can be explained by the minor impact of $B$ on $V$ compared to $A$ for low values of $\sigma_l^2$.
However, as $\sigma_l^2$ increases, the value of $B$ grows up to the point that it becomes the predominant term.
Another remarkable results of this numerical evaluations is the decrease of $V$ with the number of training iterations.
This can be explained by the fact that the losses are positive and bounded, and that the training process aims to reduce their expected value, leading simultaneously to a decreased variance.

\subsection{Impact of the Noise Variance on the BLER} \label{subsec:impact}

\begin{figure}[!t]
\begin{tikzpicture}
	  \begin{axis}[
	  compat=newest,
	  	grid=both,
	  	grid style={line width=.1pt, draw=gray!10},
	    major grid style={line width=.2pt,draw=gray!50},
	    minor tick num=0,
	    xmode=log,
	    ymode=log,
	    yticklabels={5$\cdot 10^{-4}$,6$\cdot 10^{-4}$,7$\cdot 10^{-4}$,8$\cdot 10^{-4}$,9$\cdot 10^{-4}$, 1$\cdot 10^{-3}$},
  		ytick={5e-4,6e-4,7e-4,8e-4,9e-4},
	    ymin=5e-4,
	    ymax=1e-3,
		tick scale binop= \times,	    
		ymajorgrids,
	    xlabel={$\sigma_l^2$ (MSE)},
	    ylabel={BLER},
	    	legend style={at={(0.03, 0.77)},anchor=south west},
	    	legend cell align={left},
	    		  ]
	    \addplot[blue, mark=diamond] table [x=MSE, y=nn, col sep=comma] {figs/BLER_MSE.csv};
	    \addplot[orange,  mark=o] table [x=MSE, y=alternating, col sep=comma] {figs/BLER_MSE.csv};;
	  \addlegendentry{Noisy feedback}
	  \addlegendentry{Perfect feedback}
		\end{axis}
		\end{tikzpicture}
		\vspace{-10pt}
	    \caption{Comparison of the \gls{BLER} achieved by communication systems trained with noisy and perfect feedback}
	    \label{fig:BLER_MSE}

\end{figure}
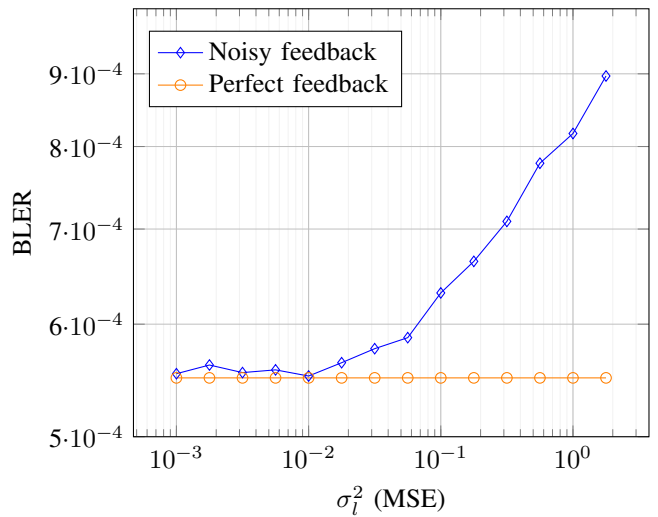

Fig.~\ref{fig:BLER_MSE} compares the \gls{BLER}, i.e., ${\Pr(\hat{m}\neq m)}$, achieved by a fully trained autoencoder using alternative training with  perfect feedback against that achieved with noisy feedback.
Both training and evaluation were done on an \gls{AWGN} channel with a \gls{SNR} of $10~\si{dB}$ as defined in (\ref{eq:snr}) in Section~\ref{sec:eval}.
Up to values of $\sigma_l^2\approx10^{-2}$, the system with noisy feedback achieves the same performance as if it was trained with a perfect feedback link.
For higher values of $\sigma_l^2$, the achieved \gls{BLER} increases quickly with $\sigma_l^2$.
These results can be explained by the high variance of the gradient loss estimator for values of $\sigma_l^2$ higher than $10^{-2}$, leading to poor training of the transmitter.

\begin{figure}[t]
\centering
  \includegraphics[width=\linewidth]{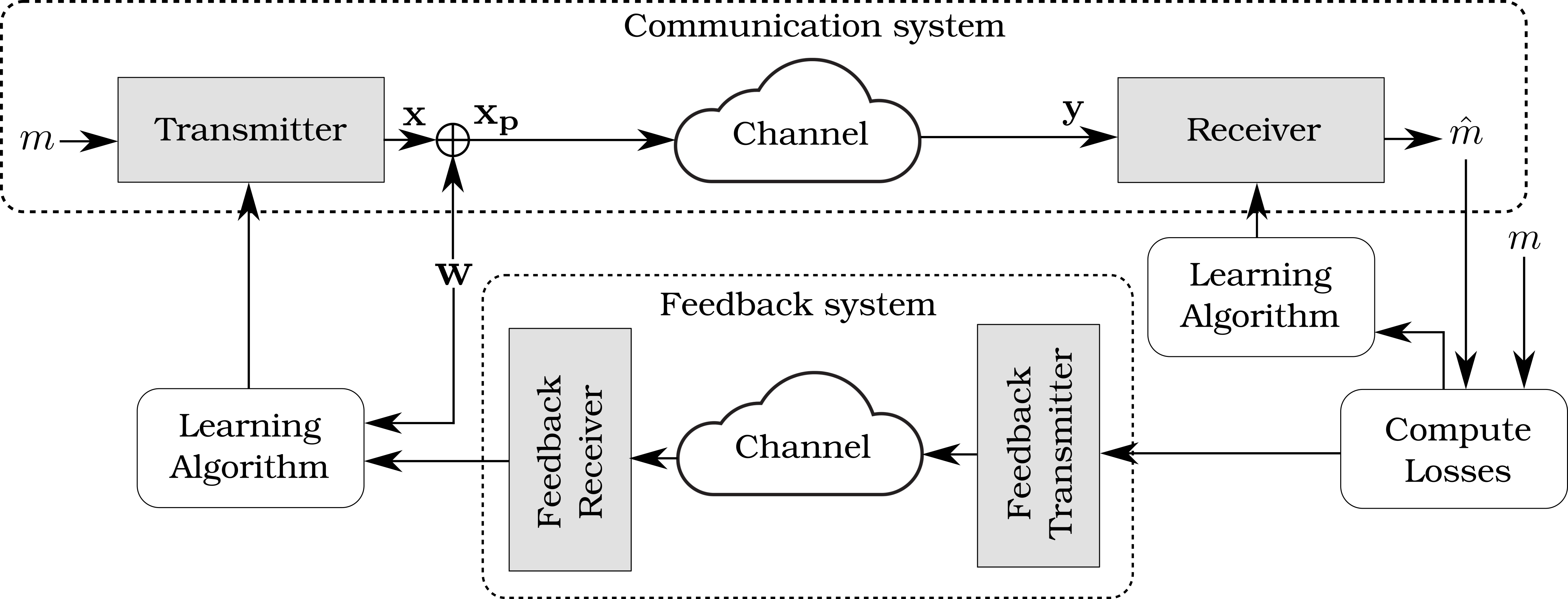}
  \caption{Training of a communication system using a feedback system}
\label{fig:full_system}
\vspace{-10pt}
\end{figure}

These results suggest that training an autoencoder with noisy feedback can lead to the same \gls{BLER} performance as with perfect feedback, as long as the \gls{MSE} of the noisy feedback link stays below a threshold, estimated here as being approximately $10^{-2}$.
Motivated by this observation, we present in the next section a system which learns to transmit real scalars. 
This system achieves a low enough \gls{MSE} so that it can be used, once trained, to feed back the losses in the alternating training scheme, as shown in Fig.~\ref{fig:full_system}, and enables the same \gls{BLER} performance as in \cite{Aoudia2018EndtoEndLO} with a perfect feedback.

\section{Learning the communication of real numbers} \label{sec:algo}

\begin{figure}[t]
\centering
  \includegraphics[width=\linewidth]{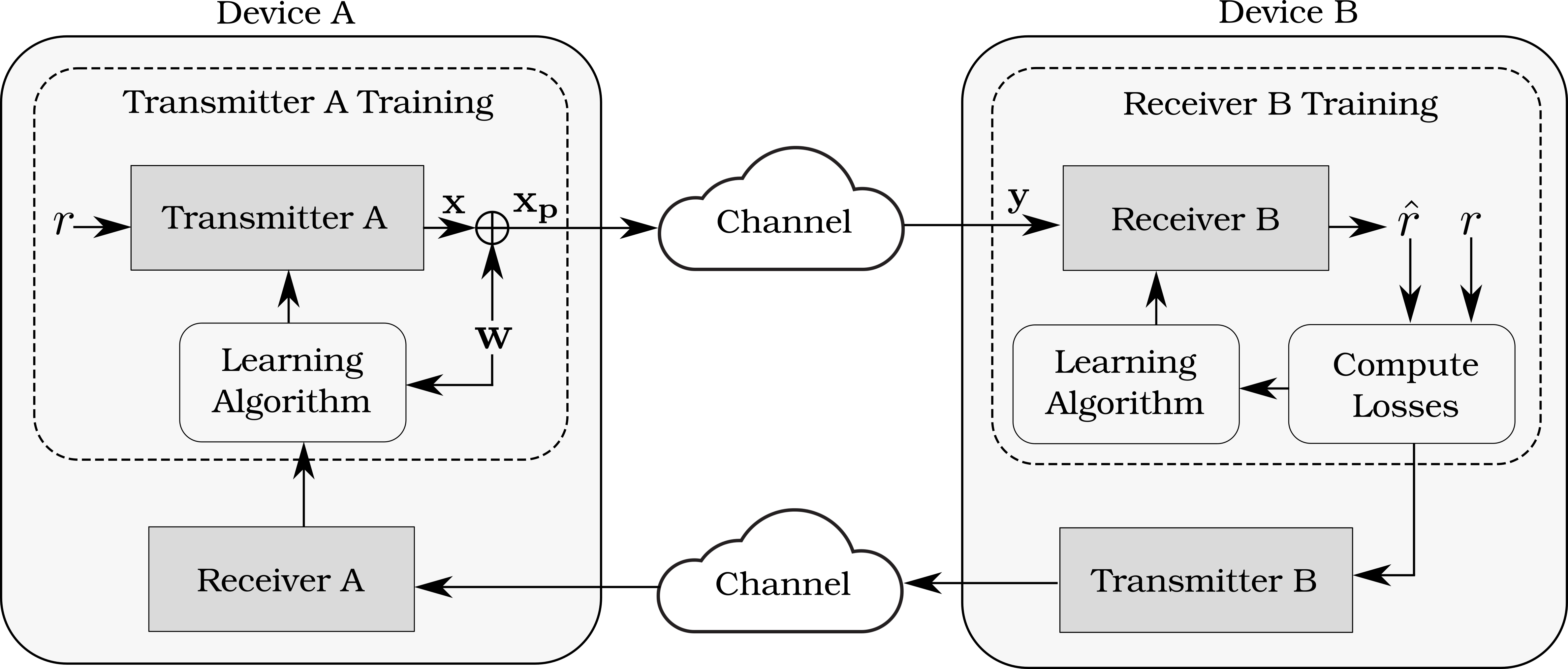}
  \caption{Training of a feedback system}
\label{fig:feedback_plot}
\vspace{10pt}
\end{figure}

This section presents an \gls{ML}-based end-to-end system for the transmission of real numbers, referred to as \textit{feedback system}, in opposition to the \textit{communication system} presented in Section \ref{sec:comm_system} which sends messages. 
The corresponding training algorithm is also introduced, and does not require any channel model or preexisting feedback link.
It is assumed that the real numbers to be transmitted take values in $[0,1]$.
Two \emph{devices}, denoted by A and B, are each assumed to be equipped with a transmitter and a receiver, allowing data transmission and reception. Transmitter A aims to  communicate real numbers to receiver B, and similarly transmitter B aims to  communicate real numbers to receiver A. This is illustrated in Fig.~\ref{fig:feedback_plot}.

Transmitter $X \in \{\text{A}, \text{B}\}$ implements the mapping: $ f^{T_X}_{\thetav_{T_X}} : \RR \to \CC^{N_f}$, where $\thetav_{T_X}$ is the parameter vector of transmitter $X$, and $N_f$ the number of channel uses required to send a single real number.
The final layer normalizes the average energy per channel symbol, such that $\EE \LSB \frac{1}{N_{f}} \lVert\xv \rVert^2_2 \RSB=1$.
Receiver $X \in \{\text{A}, \text{B}\}$ implements the mapping $f^{R_X}_{\thetav_{R_X}} : \CC^{N_f} \to \RR$, $\thetav_{R_X}$ being the parameter vector of receiver $X$.
The proposed training algorithm, which can be seen as an extension of the one proposed in \cite{Aoudia2018EndtoEndLO}, is presented in the following.

\subsection{Receiver Training}\label{subsec:rxtrain}

Due to the symmetry between the devices A and B, only training of receiver B is described.
First, transmitter A generates a minibatch of real numbers of size $S_f$ containing realizations $\rv$ distributed in $[0,1]$.
Each of these realizations is encoded by the transmitter into $N_f$ complex symbols, creating an $S_f$--by--$N_f$ matrix $\Xm$ that is sent over the channel (row-by-row).
The receiver receives the perturbed symbols $\Ym$ and decodes them into real numbers $\hat{\rv}$.
Finally, \gls{SGD} is performed on $f^{R_B}_{\thetav_{R_B}}$ to minimize the \gls{MSE} between $\rv$ and $\hat{\rv}$, the \gls{CE} only being used when transmitting messages.
It is assumed that device B is aware of the true real numbers sent by the device A.
This can easily be done in practice, e.g., using pseudo-random number generators initialized with the same seed.
Algorithm~\ref{alg:receiver_training} shows training of a receiver.

\begin{algorithm}[t]
\caption{ \textbf{ : \hspace{1pt} function} \sc TrainReceiverB}
\label{alg:receiver_training}
\begin{algorithmic}[1]
		\Repeat
        \State $\rv \gets \Call{TrainingSource}{S_f}$ \tikzmark{top1}\tikzmark{right1}
        \State $\Xm \gets f^{(T_A)}_{\thetav_{T_A}}(\rv)$ \tikzmark{bottom1}
        \State \LineComment{Channel : $\Ym \sim P(\Ym|\Xm)$}
        \State $\hat{\rv} \gets f^{(R_B)}_{\thetav_{R_B}}(\Ym)$ \tikzmark{top2}
        \State $\rv \gets \Call{TrainingSource}{S_f}$ \tikzmark{right2}
        \State $\Call{SGD}{\thetav_{R_B}, \rv, \hat{\rv}}$ \tikzmark{bottom2}
        \Until{Stop criterion is met}
\end{algorithmic}
\AddNote{top1}{bottom1}{right1}{Transmitter A}
\AddNote{top2}{bottom2}{right2}{Receiver B}
\vspace{-5pt}
\end{algorithm}

\subsection{Transmitter Training}\label{subsec:txtrain}

Similarly, only training of transmitter A is presented due to symmetry.
Transmitter A starts by creating a minibatch of real numbers of size $S_f$ and encodes each example into $N_f$ complex symbols to form the matrix $\Xm$ of size $S_f$--by--$N_f$.
Note that the batch size of the receiver and transmitter training are chosen equal for simplicity but could be different.
To enable exploration, a stochastic perturbation is added to the output of the transmitter. More precisely, a perturbation matrix $\Wm$ is generated by sampling a circular complex Gaussian distribution with variance $\sigma_{f}^2$.
In order to achieve unit energy channel symbols, the output of the transmitters  are scaled before the perturbation is added, and the matrix containing the sent symbols is given by $\Xm_p = \sqrt{1 - \sigma_{f}^2 } \Xm + \Wm$.

The receiver receives the matrix $\Ym$ and decodes the symbols into real numbers $\hat{\rv}$. The per-examples losses are computed as being the per-symbol squared error.
Because the system is trained to communicate real numbers in both ways, it is possible to transmit the losses back to the device A using the transmitter B, and thus avoiding the need of a preexisting reliable feedback link.
Finally, \gls{SGD} is performed on the transmitter weights using an estimation of the gradient obtained by the policy gradient theorem~\cite{NIPS1999_1713}.
Algorithm~\ref{alg:transmitter_training} summarizes the training procedure of a transmitter.

\begin{algorithm}[h]
\caption{\textbf{ : \hspace{1pt} function} \sc TrainTransmitterA}
\label{alg:transmitter_training}
\begin{algorithmic}[1]
		\Repeat
    		\State $\rv \gets \Call{TrainingSource}{S_f}$ \tikzmark{top1} \tikzmark{right1}
    		\State $\Xm \gets f^{(T_A)}_{\thetav_{T_A}}(\rv)$
    		\State $\Wm \gets \Call{Perturbation}{\sigma_{f}}$ 
    		\State $\Xm_p \gets \sqrt{1 - \sigma_{f}^2 } \Xm + \Wm$\tikzmark{bottom1}
   		\State \LineComment{Channel : $\Ym \sim P(\Ym|\Xm_p)$ }
    		\State $\hat{\rv} \gets f^{(R_B)}_{\thetav_{R_B}}(\Ym)$ \tikzmark{top2}
    		\State $\rv \gets \Call{TrainingSource}{S_f}$ 
    		\State $\lv \gets \Call{PerExampleLosses}{\rv, \hat{\rv}}$ \tikzmark{right2}
    		\State $\Call{SendWithTransmitterB}{\lv}$ \tikzmark{bottom2}
    		\State $\hat{\lv} \gets \Call{ReceiveWithReceiverA}{}$ \tikzmark{top3}\tikzmark{right3}
   		\State $\Call{SGD}{\thetav_{T_A}, \hat{\lv}, \Wm}$\tikzmark{bottom3}
   		\Until{Stop criterion is met}
\end{algorithmic}
\AddNote{top1}{bottom1}{right1}{Transmitter A}
\AddNote{top2}{bottom2}{right2}{Receiver B}
\AddNote{top3}{bottom3}{right3}{Transmitter A}
\vspace{-5pt}
\end{algorithm}

\subsection{Main Loop}
\label{sec:training}

Algorithm~\ref{alg:main_loop} shows the main loop of the proposed algorithm.
Each iteration is divided into two steps.
In the first step, transmitter A and receiver B are trained iteratively until a certain criterion is met.
In the second step, transmitter B and receiver A are trained in the same way.
The intuition is that, by alternating between optimization of a transmitter and a receiver, communication errors will decrease.
A key idea of this algorithm is to use one transmitter-receiver pair to transmit the losses which are needed for training of the other pair. We expect that as the training progresses, each pair will get better at reliably communicating real numbers (and hence losses), enabling more accurate optimization of the other pair.

\begin{algorithm}[h!]
\caption{\textbf{ : \hspace{1pt}} Main loop}
\label{alg:main_loop}
\begin{algorithmic}[1]
\While {\textrm{Stop criterion not met}}
  \While {\textrm{Stop criterion not met}}
      \State \Call{TrainTransmitterA}{}
      \State \Call{TrainReceiverB}{}
    \EndWhile
  \While {\textrm{Stop criterion not met}}
      \State \Call{TrainTransmitterB}{}
      \State \Call{TrainReceiverA}{}
    \EndWhile
\EndWhile
\end{algorithmic}
\vspace{5pt}
\end{algorithm}

\section{Evaluation Results}
\label{sec:eval}

\begin{figure*}
    \centering
    \begin{subfigure}{0.45\linewidth}
		\begin{tikzpicture}
			\begin{axis}[
				ymode=log,
				grid=both,
				grid style={line width=.1pt, draw=gray!10},
				major grid style={line width=.2pt,draw=gray!50},
				minor tick num=3,
				xlabel={SNR (dB)},
				ylabel={MSE},
				legend style={at={(0.03, 0.03)},anchor=south west},
				xtick={-4, 0, 4, 8, 12, 16},
				legend cell align={left},
			]
				\addplot[blue, mark=diamond] table [x=SNR, y=RL, col sep=comma] {figs/MSE_SNR_RL_Ana.csv};
				\addplot[red, mark=triangle] table [x=SNR, y=Ana, col sep=comma] {figs/MSE_SNR_RL_Ana.csv};
				\addlegendentry{Feedback system}
				\addlegendentry{Analog transmission}
			\end{axis}
		\end{tikzpicture}
		\caption{\gls{AWGN} channel}
		\label{fig:MSE_SNR_AWGN}
	 \end{subfigure} \qquad
    \begin{subfigure}{0.45\linewidth}
	    \begin{tikzpicture}
		  \begin{axis}[
		    ymode=log,
		    grid=both,
		    grid style={line width=.1pt, draw=gray!10},
		    major grid style={line width=.2pt,draw=gray!50},
		    minor tick num = 9,
		    xlabel={SNR (dB)},
		    ylabel={MSE},
		    legend style={at={(0.03, 0.03)},anchor=south west},
		    legend cell align={left},
		  ]
			\addplot[blue, mark=diamond] table [x=SNR, y=RL, col sep=comma] {figs/MSE_SNR_rayleigh.csv};
			\addplot[red, mark = triangle] table [x=SNR, y=Ana, col sep=comma] {figs/MSE_SNR_rayleigh.csv};

			\addlegendentry{Feedback system}
			\addlegendentry{Analog transmission}
			\end{axis}    
		\end{tikzpicture}
	    \caption{\gls{RBF} channel}
	    \label{fig:MSE_SNR_RBF}
	\end{subfigure}

	\caption{\gls{MSE} achieved by the proposed feedback system and an analog transmission system over \gls{AWGN} and \gls{RBF} channels}
	\label{fig:feedback_sys_eval}
\vspace{-5pt}
\end{figure*}
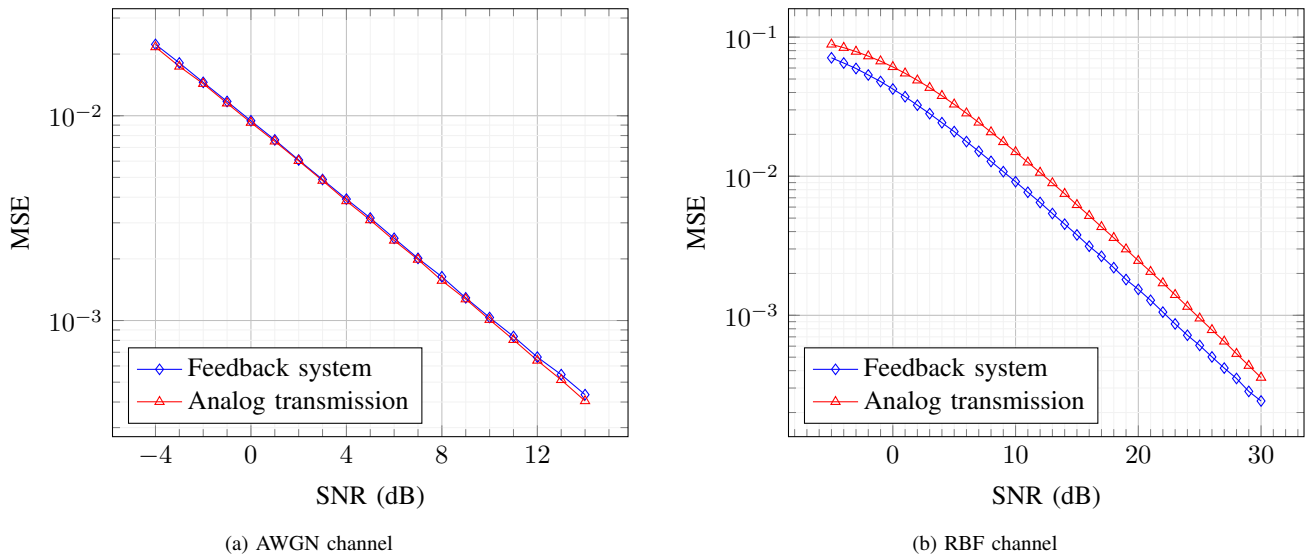

We first evaluate numerically the performance of the feedback system described above.
Then, we compare the \glspl{BLER} of the communication system of Section~\ref{sec:comm_system} trained using the alternating scheme with (i) perfect feedback~\cite{Aoudia2018EndtoEndLO} and (ii) the proposed feedback system. The \gls{SNR} is defined as
\begin{equation}\label{eq:snr}
  \text{SNR} = \frac{ \EE \LSB \frac{1}{N_{c}} \lVert\xv \rVert^2_2 \RSB }{\sigma^2} = \frac{1}{\sigma^2}
\end{equation}
where the last equality is due to normalization of channel symbols, and $\sigma^2$ is the channel noise variance.
Training was performed using the Adam~\cite{Kingma15} optimizer. Evaluation was done on \gls{AWGN} and \gls{RBF} channels. For the feedback system, only the results for one transmitter-receiver pair are presented, as they are similar for the other one.

\subsection{Feedback System Transmitters and Receivers Architectures}

Both transmitters of the feedback system are made of two dense layers, a first one of $10N_f$ units with ELU activation functions~\cite{Goodfellow-et-al-2016-Book}, and a second one of $2N_f$ units with linear activations.
The output of the second layer form the real and imaginary parts of the $N_f$ complex channel symbols used to transmit a real number.
A last layer normalizes the average energy per symbol.

Regarding the receivers, their first layers split the $N_f$ received complex symbols into $2N_f$ real numbers. Then, different architectures are used for the \gls{AWGN} and \gls{RBF} channel.
In the \gls{AWGN} channel case, a single hidden dense layer with $10N_f$ units and ReLu activations~\cite{Goodfellow-et-al-2016-Book}, followed by an output layer consisting of a single-neuron dense layer with linear activation was found to be enough.
In the \gls{RBF} channel case, the receiver adopts a \gls{RTN}~\cite{8054694} architecture, similar to the one used in~\cite{Aoudia2018EndtoEndLO}, the only differences being that the second to last layer has $10N_f$ units, and that the last layer has a single unit with linear activation.

\subsection{Evaluation of the Feedback System}

In Fig.~\ref{fig:feedback_sys_eval}, the \gls{MSE} achieved by the proposed feedback system and by an analog scheme are presented for \gls{AWGN} and \gls{RBF} channels.
Considering the \gls{AWGN} and \gls{RBF} case, $N_f$ was set to $4$ and $5$ respectively, and training was performed at an \gls{SNR} of $10\:$\si{dB} and $20\:$\si{dB} respectively.
 $\sigma_{f}^2$ was set to $0.02$, $S_{f}$ to $10^{5}$, and real numbers $r$ to be sent are drawn from a uniform distribution on [0,1], i.e., $r \thicksim \Uc(0,1)$.

In the \gls{AWGN} channel case, each real number was transmitted using $N_f = 4$ channel uses.
Before transmitting real numbers with the analog system, they are centered and scaled to have zero-mean and unit variance.
The transmitted channel symbols vector corresponding to a real number $r$ is therefore
\begin{equation}\label{eq:analog_transmitter_awgn}
\xv = \frac {r - \EE \LSB r \RSB  + j \LB r - \EE \LSB r \RSB \RB}{\sqrt{\text{2var}(r)}}{{1}}_{N_f}.
\end{equation}
Decoding is done by averaging over the repetitions and proper scaling:
\begin{equation}\label{eq:analog_receiver_awgn}
\resizebox{0.88\hsize}{!}{$
\hat{r} = \LSB \frac{\sqrt{\text{2var}(r)}}{2N_f} {\sum}_{j = 1}^{N_f} \LB \Re\LB y^{(j)} \RB+\Im\LB y^{(j)} \RB \RB + \EE\LSB r \RSB \RSB_{\top0}^{\bot1}.
$}
\end{equation}
Fig.~\ref{fig:MSE_SNR_AWGN} shows that over an \gls{AWGN} channel, the feedback system and the analog system achieve similar performance.
Moreover, when the \gls{SNR} is greater than $0\:$\si{dB}, the \gls{MSE} is lower than $10^{-2}$, meaning from our previous analysis (see Section~\ref{sec:effect_noisy_losses}) that either the feedback system or an analog system with repetition can be used in lieu of a perfect feedback link to perform alternating training.

Regarding the \gls{RBF} channel, comparison was done to an analog scheme with repetition over four channel uses in Fig.~\ref{fig:MSE_SNR_RBF}.
A pilot was added to perform equalization.
Similarly to the \gls{AWGN} case, real numbers are centered and scaled to have zero-mean and unit variance before sending.
Detection is done by first computing an estimate of the channel gain $\hat{h}$ using the pilot, before averaging over the equalized received symbols and scaling:
\begin{multline}\label{eq:analog_receiver_awgn}
\resizebox{0.88\hsize}{!}{$
\hat{r} = \LSB \frac{\sqrt{2\text{var}(r)}}{2(N_f-1)} {\sum}_{j = 1}^{N_f} \LB \Re\LB\frac{y^{(j)}}{\hat{h}}\RB+\Im\LB\frac{y^{(j)}}{\hat{h}}\RB \RB + \EE\LSB r \RSB \RSB_{\top0}^{\bot1}.
$}
\end{multline}

For fairness, the number of channel uses for the feedback system was set to $N_f = 5$.
Fig.~\ref{fig:MSE_SNR_RBF} reveals that the feedback system outperforms the analog approach, with a maximum gain of approximately $3\si{dB}$ being achieved around an \gls{MSE} of $10^{-2}$.
On an \gls{RBF} channel, these results show that the feedback system is capable of providing an \gls{MSE} of $10^{-2}$ at $\text{SNR} = 10~\si{dB}$, which should be sufficient for the alternative training.

\begin{figure*}[h!]
    \centering
    \begin{subfigure}{0.45\linewidth}
		\begin{tikzpicture}
			  \begin{axis}[
			    ymode=log,
			    grid=both,
			    grid style={line width=.1pt, draw=gray!10},
			    major grid style={line width=.2pt,draw=gray!50},
			    minor tick num=3,
			    xlabel={SNR(\si{dB})},
			    ylabel={BLER},
			    	legend style={at={(0.03, 0.03)},anchor=south west},
			    xtick={-4, 0, 4, 8, 12, 16},
			    legend cell align={left},
			  ]
				\addplot[blue, mark=diamond] table [x=SNR, y=nn, col sep=comma] {figs/BLER_SNR_AWGN.csv};
				\addplot[orange, mark=o] table [x=SNR, y=auto, col sep=comma] {figs/BLER_SNR_AWGN.csv};
				\addplot[black, dashed] table [x=SNR, y=QPSK, col sep=comma] {figs/BLER_SNR_AWGN.csv};
				\addplot[black] table [x=SNR, y=Agrell, col sep=comma] {figs/BLER_SNR_AWGN.csv};
				\addlegendentry{Autoenc. / feedback system}
				\addlegendentry{Autoenc. / perfect feedback}
				\addlegendentry{QPSK}
				\addlegendentry{Agrell}
			\end{axis}
		\end{tikzpicture}
		\caption{\gls{AWGN} channel}
		\label{fig:BLER_SNR_AWGN}
	\end{subfigure} \qquad
	\begin{subfigure}{0.45\linewidth}
		\begin{tikzpicture}
		\begin{axis}[
			ymode=log,
			grid=both,
			grid style={line width=.1pt, draw=gray!10},
			major grid style={line width=.2pt,draw=gray!50},
			minor tick num=9,
			xlabel={SNR(\si{dB})},
			ylabel={BLER},
			ymin = 4e-4,
			legend style={at={(0.03, 0.03)},anchor=south west},
			legend cell align={left},
		]
			\addplot[blue, mark=diamond] table [x=SNR, y=feedback, col sep=comma] {figs/BLER_SNR_rayleigh.csv};
			\addplot[orange, mark=o] table [x=SNR, y=perfect, col sep=comma] {figs/BLER_SNR_rayleigh.csv};
			\addplot[black, dashed] table [x=SNR, y=QPSK, col sep=comma] {figs/BLER_SNR_rayleigh.csv};
			\addplot[black] table [x=SNR, y=Agrell, col sep=comma] {figs/BLER_SNR_rayleigh.csv};
			\addlegendentry{Autoenc. / feedback system}
			\addlegendentry{Autoenc. / perfect feedback}
			\addlegendentry{QPSK}
			\addlegendentry{Agrell}
		\end{axis}
		\end{tikzpicture}
		\caption{\gls{RBF} channel}
		\label{fig:BLER_SNR_RBF}
	\end{subfigure}
	\caption{\gls{BLER} achieved by the compared approaches over \gls{AWGN} and \gls{RBF} channels}
	\label{fig:princ_sys_eval}
\end{figure*}
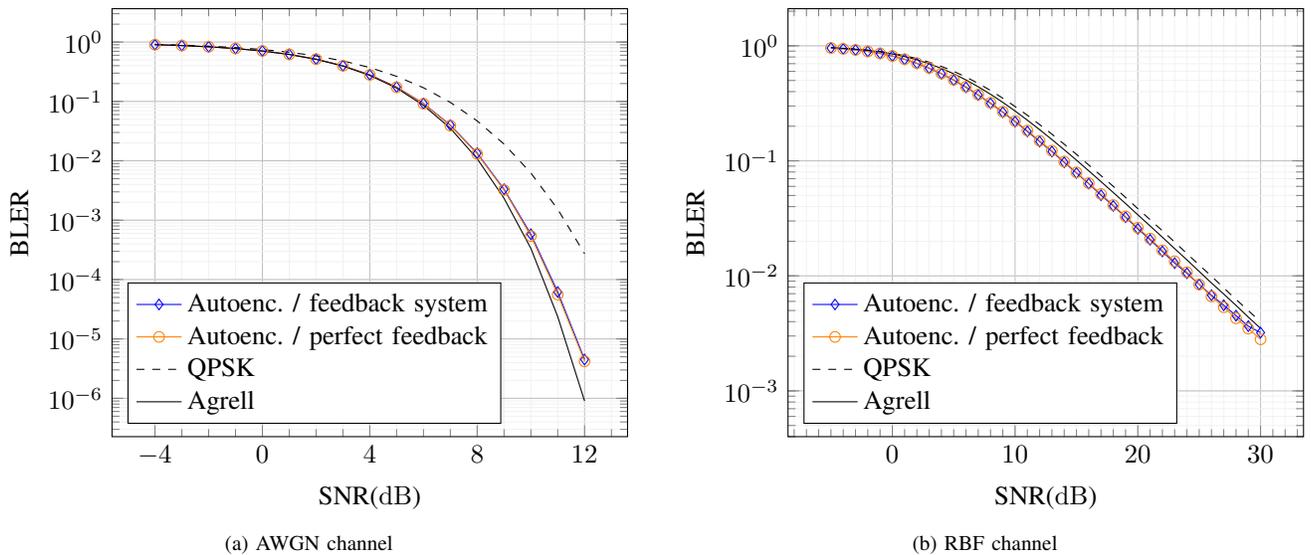
\vspace{-5pt}
\subsection{Alternating Training using the Feedback System}

Next, the performance of communication systems as introduced in Section~\ref{sec:comm_system} are evaluated.
We are interested in \glspl{BLER} achieved by communication systems, as introduced in Section \ref{sec:comm_system}, when trained using feedback systems instead of a perfect feedback link. This is illustrated in Fig.~\ref{fig:full_system}.
The communication systems transmit messages from a set of $M = 256$ messages, uses the same architecture as in~\cite{Aoudia2018EndtoEndLO}, and losses are clipped in $[0,1]$ before being sent back to the transmitter to match feedback systems' training conditions.

Comparison is done to a system trained with perfect link, as well as to \gls{QPSK} and Agrell~\cite{Agrell16} modulation schemes.
Agrell is a subset of the E8 lattice, designed by numerical optimization to solve the sphere packing problem for $M = 256$.
To enable \gls{QPSK} and Agrell on an \gls{RBF} channel, an additional pilot symbol was used to perform equalization.
For the \gls{AWGN} and \gls{RBF} channel, the communication and feedback systems were trained at $10\:$\si{dB} and $20\:$\si{dB} respectively in both directions, with $\sigma^2_{c} = \sigma^2_{f} = 0.02$ and $S_c = S_f = 10^5$.
The number of channel uses was set to $N_c = N_f = 4$ for the \gls{AWGN} channel, and $Nc = N_f = 5$ for fairness for the \gls{RBF} channel.

Fig.~\ref{fig:princ_sys_eval} shows the \gls{BLER} of trained communication systems over \gls{AWGN} and \gls{RBF} channels.
It can be seen that the communication systems trained using the feedback system achieve the same performance as those trained with perfect feedback. This demonstrates the effectiveness of the proposed approach.
In the \gls{AWGN} channel case, our communication system outperforms \gls{QPSK} but has a slightly higher \gls{BLER} than Agrell.
This is not surprising as Agrell is a close-to-optimal solution to the sphere packing problem, leading to high performance on the \gls{AWGN} channel.
However, our communication system outperforms both \gls{QPSK} and Agrell over the \gls{RBF} channel. This shows that it is able to discover a more robust scheme, even without any channel model or reliable feedback link.

\section{Conclusion}

We have shown that alternating training~\cite{Aoudia2018EndtoEndLO} of \gls{ML}-based communication systems can be performed with noisy feedback, up to a certain \gls{MSE}, without performance loss.
We then proposed a feedback system able to learn the transmission of real numbers without channel model or preexisting feedback link.
This feedback system can be used in lieu of the perfect feedback link to perform alternating training.
Finally, evaluations show that the feedback system leads to identical performance compared to that achieved with a perfect feedback link, conditioned on a sufficiently high but realistic training \gls{SNR}. Moreover, our communication system outperforms both QPSK and a highly-optimized higher-order modulation scheme on an \gls{RBF} channel.

\bibliographystyle{IEEEtran}
\bibliography{IEEEabrv,bibliography}

\end{document}